# Fermi Velocity Renormalization in Graphene Probed by Terahertz Time-Domain Spectroscopy

**Patrick R. Whelan,**[1,2,*] **Qian Shen,**[3,4] **Binbin Zhou,**[3] **I. G. Serrano,**[5] **M. Venkata-Kamalakar,**[5] **David M. A. Mackenzie,**[6] **Jie Ji,**[1,2] **Deping Huang,**[7] **Haofei Shi,**[7] **Da Luo,**[8] **Meihui Wang,**[9] **Rodney S. Ruoff,**[8,9,10,11] **Antti-Pekka Jauho**[1,2]**, Peter U. Jepsen,**[2,3] **Peter Bøggild**[1,2] **and José M. Caridad**[1,2,*]

[1]*DTU Physics, Technical University of Denmark, Ørsteds Plads 345C, DK-2800 Kongens Lyngby, Denmark*
[2]*Center for Nanostructured Graphene (CNG), Technical University of Denmark, Ørsteds Plads 345C, DK-2800 Kongens Lyngby, Denmark*
[3]*DTU Fotonik, Technical University of Denmark, Ørsteds Plads 343, DK-2800 Kongens Lyngby, Denmark*
[4]*School of Information Engineering, Nanchang University, Nanchang 330031, P. R. China*
[5]*Department of Physics and Astronomy, Uppsala University, Box 516, SE 751 20, Uppsala, Sweden*
[6]*Department of Electronics and Nanoengineering, Aalto University, P.O. Box 13500, FI-00076 Aalto, Finland*
[7]*Chongqing Institute of Green and Intelligent Technology, Chinese Academy of Sciences, 266 Fang Zheng Ave., Chongqing 400714, P. R. China*
[8]*Center for Multidimensional Carbon Materials (CMCM), Institute for Basic Science (IBS), Ulsan 44919, Republic of Korea*
[9]*Department of Chemistry,* [10]*School of Materials Science and Engineering, and* [11]*School of Energy and Chemical Engineering, Ulsan National Institute of Science and Technology (UNIST), Ulsan 44919, Republic of Korea*

Emails: patwhe@dtu.dk, jcar@dtu.dk

**Abstract**

We demonstrate terahertz time-domain spectroscopy (THz-TDS) to be an accurate, rapid and scalable method to probe the interaction-induced Fermi velocity renormalization $v_F^*$ of charge carriers in graphene. This allows the quantitative extraction of all electrical parameters (DC conductivity $\sigma_{DC}$, carrier density $n$, and carrier mobility $\mu$) of large-scale graphene films placed on arbitrary substrates via THz-TDS. Particularly relevant are substrates with low relative permittivity (< 5) such as polymeric films, where notable renormalization effects are observed even at relatively large carrier densities (> $10^{12}$ cm$^{-2}$, Fermi level > 0.1 eV). From an application point of view, the ability to rapidly and non-destructively quantify and map the electrical ($\sigma_{DC}$, $n$, $\mu$) and electronic ($v_F^*$) properties of large-scale graphene on generic substrates is key to utilize this material in applications such as metrology, flexible electronics as well as to monitor graphene transfers using polymers as handling layers.

## 1. Introduction

In recent years, numerous experimental studies[1–6] have revealed many-body interactions to play an unexpectedly important role in graphene. Among other effects,[7–10] long-range electron-electron interactions in graphene induce a momentum-dependent renormalization of the Fermi velocity of charge carriers in the monolayer. Such corrections are expected to be notable close to the charge neutrality point where the density of states vanishes.[7–10] Moreover, these interactions additionally depend on the dielectric environment surrounding the graphene layer.[7–10] Indeed, the need to consider a renormalized Fermi velocity of charge carriers in graphene has already been demonstrated in different experiments undertaken in high quality graphene samples at low carrier densities $n < 10^{11}$ cm$^{-2}$.[1–5] Such experiments include electrical transport[2,3] and plasmon propagation[5] in the monolayer as well as angle-resolved photoemission spectroscopy[1] and infrared spectromicroscopy[4] measurements.

In the present study, we demonstrate terahertz time-domain spectroscopy (THz-TDS) to be a sensitive, rapid and scalable method to probe and spatially map the renormalized Fermi velocity $v_F^*$ of charge carriers in graphene placed on arbitrary substrates. This is done by taking into account a recently developed Hartree-Fock theory[9] (able to account for the effect of interacting electrons in graphene without the need for fitting parameters) when extracting the electrical properties of graphene films from THz spectra. We verify the validity of this approach by showing an excellent quantitative agreement between the electrical parameters (DC conductivity $\sigma_{DC}$, scattering time $\tau$, carrier density $n$ and mobility $\mu$) extracted via THz-TDS conductivity spectra (accounting for $v_F^*$) with those obtained from traditional metal contact-based Hall measurements ($n_H$ and $\mu_H$) of graphene on different substrates. From a more applied point of view, we show that Fermi velocity renormalization effects must be taken into account to accurately interpret THz spectra and obtain the electrical properties of graphene placed on substrates with low permittivity ($\varepsilon < 5$) even at relatively high charge carrier densities $n > 10^{12}$ cm$^{-2}$. Examples of such substrates include thin polymeric films, highly relevant for applications in flexible electronics.[11–14] Furthermore, we demonstrate that our results are general and consistent with previous results reported for substrates with higher permittivity.[15,16] In other words, the methodology reported here enables to rapidly quantify and map the electrical ($\sigma_{DC}$, $\tau$, $n$, $\mu$) and electronic ($v_F^*$) properties of graphene films placed on arbitrary substrates via THz-TDS.

The manuscript is organized as follows. In section 2, we point out the need to renormalize the Fermi velocity of graphene charge carriers $v_F^*$ in order to obtain the key electrical parameters ($\sigma_{DC}$, $\tau$, $n$, $\mu$) of graphene films placed on arbitrary substrates via THz-TDS measurements. In section 3, we describe our experimental methods. In section 4, we demonstrate the accurate extraction of $\sigma_{DC}$, $\tau$, $n$, $\mu$ and $v_F^*$ of large-scale graphene placed on two flexible polymer substrates (polyethylene naphthalate, PEN, and polyethylene terephthalate, PET) with low dielectric permittivity (< 5) via THz-TDS. In particular, we show that when accounting for a renormalized $v_F^*$, the electrical parameters extracted via THz-TDS are in good quantitative agreement with those obtained via electrical Hall measurements. Furthermore, for completeness, in Section 5 we discuss and demonstrate *i)* the validity of this procedure to extract the electrical parameters of graphene on rigid, commonly used low ε substrates such as $SiO_2$ and also *ii)* show the consistency of the here presented procedure with previously reported measurements of graphene on substrates with higher permittivities such as silicon nitride[15] and silicon carbide[16]. Finally, our conclusions are summarized in section 6.

## 2. Fermi velocity renormalization and extraction of electrical and electronic parameters in graphene placed on arbitrary substrates via THz-TDS

Interacting electrons in graphene behave as independent quasiparticles with a renormalized energy dispersion, which results in different corrections to the electronic and optical properties of the monolayer.[1,7–9] As demonstrated here, some of these corrections can be accurately probed via THz-TDS. The frequency-dependent sheet conductivity of graphene, $\sigma_s(\omega)$ can be efficiently extracted via THz-TDS measurements (see Methods) due to the fact that THz transients are particularly sensitive to absorption by free carriers.[17,18] In the local approximation,[5] this conductivity is well described by a Drude response, yielding[17,19–21]

$$\sigma_s(\omega) = \frac{\sigma_{DC}}{1-i\omega\tau} \qquad (1)$$

where $\sigma_{DC}$ is related to the Drude weight $D_W$ and the scattering rate $\Gamma = \tau^{-1}$ as $\sigma_{DC} = D_W/\pi\Gamma$.

As such, by fitting the Drude model to the real part of each recorded conductivity spectrum, $\sigma_{DC}$ and $\tau$ can be spatially mapped in graphene as already reported in literature.[15,17,22,23] Then, in the limit of thermal energies being smaller than the Fermi level of graphene charge carriers[24] ($k_B T \ll \varepsilon_F$, Fermi-liquid regime[25]), $D_W \approx (e^2/\hbar^2)\varepsilon_F$ and one can extract the remaining electrical properties n and μ by taking into account the relations $\sigma_{DC} = e^2 v_F \tau \sqrt{n}/(\hbar\sqrt{\pi})$[15,17,26] and $\sigma_{DC} = \mu n e$:

$$n = \frac{\pi\hbar^2}{e^4 v_F^2}\left(\frac{\sigma_{DC}}{\tau}\right)^2 \qquad (2)$$

$$\mu = \frac{e^3 v_F^2}{\pi\hbar^2}\frac{\tau^2}{\sigma_{DC}} \qquad (3)$$

where $v_F$ is the Fermi velocity of graphene's charge carriers, $\hbar$ is the reduced Planck's constant, and $e$ is the elementary charge. Equations 2 and 3 show that the extraction of both *n* and *μ* via THz-TDS is highly sensitive to (the square of) $v_F$, parameter which is assumed to be constant ~$1\cdot10^6$ m/s in previous studies [15-17]. However, this approximation is only valid to a first-order: as anticipated, interacting electrons induce a momentum-dependent renormalization of the Fermi velocity of charge carriers in graphene $v_F^*$.[1,7–9] As such, $v_F$ should be replaced by $v_F^*$ in Equations 2 and 3 to accurately extract $n$ and $\mu$ via THz-TDS. Furthermore, we point out that $D_W$ (thus the measured $\sigma_{DC}$, see Equation 1) is also affected by these interactions in Dirac systems such as graphene, following the same renormalization as the Fermi velocity: $D_W^*/D_W = v_F^*/v_F$.[9] For completeness, we note that interaction effects might also introduce modifications to the optical conductivity of graphene $\sigma_s(\omega)$.[9] However, such additional corrections can be disregarded at THz frequencies, since the energy scale (~4 meV) is considerably smaller than the hopping parameter of $\pi$-bands in graphene (~2.7 eV).[9]

As mentioned in the introduction, the need to introduce corrections to the Fermi velocity of charge carriers of graphene has already been reported for numerous types of graphene samples at low carrier densities ($n < 10^{11}$ cm$^{-2}$).[1–5] In these experiments, $v_F^*$ is commonly adjusted to the predicted scaling behaviour[1,7–9] by fitting the momentum cut-off parameter $\Lambda$ and/or the effective dielectric constant in the system.[2–4] Instead, in the present study, we make use of a recently developed Hartree-Fock (HF) theory able to account for both dielectric and self-screening effects without the need for fitting parameters.[9] The model has already been successfully tested in different experimental studies[9] and *i)* justifies the need for renormalization of the Fermi velocity of graphene charge carriers even at relatively large carrier densities $n > 1 \times 10^{12}$ cm$^{-2}$ in samples placed on substrates with low dielectric permittivity $\varepsilon$ as well as *ii)* enables us to obtain accurate quantitative spatial maps of $v_F^*$ as well as $n$, $\mu$ of graphene placed *on arbitrary substrates* with an unknown carrier concentration $n$ via THz-TDS.

Within this HF theory, the expression for $v_F^*$ at low energies is given by[9]

$$\frac{v_F^*}{v_F} = 1 + C(\alpha)\alpha \ln(\Lambda/k_F) \qquad (4)$$

where $\Lambda = 1.75$ Å$^{-1}$ (obtained here as a fit to the actual dispersion relation, i.e. including the presence of interacting electrons[9]), $\alpha = e^2/(4\pi\hbar v_F \varepsilon \varepsilon_0)$ is the fine-structure constant of graphene, $\varepsilon_0$ is the permittivity of vacuum. Here, $v_F$ is the bare Fermi velocity 0.85×10$^6$ m/s obtained in the LDA limit[1] where $\varepsilon \to \infty$ and $k_F = \sqrt{\pi n}$ is the Fermi momentum of the charge carriers. $C(\alpha)$ accounts for the self-screened interaction between carriers and can be accurately described via the random phase approximation (RPA) within the Dirac cone approximation as $C(\alpha) = \{4[1+(\pi/2)\alpha]\}^{-1}$.[9] Figure 1(a) shows $v_F^*$ according to equation 4 as a function of effective permittivity $\varepsilon = (\varepsilon_S+1)/2$ for a system consisting of non-encapsulated graphene placed on a substrate with relative permittivity $\varepsilon_S$ at five constant carrier densities $n$. Moreover, figure 1(b) shows $v_F^*$ for different $n$ at constant $\varepsilon_S$ instead. Both figures clearly predict an appreciable variation of $v_F^*$ with respect to the bare Fermi velocity $v_F$ when graphene is supported on substrates with $\varepsilon_S < 5$ ($\varepsilon < 3$) occurring even at large carrier densities $n > 10^{12}$ cm$^{-2}$. Figure 1(b) also shows the $v_F^*$ extracted from our experimental measurements (see below) for the given $\varepsilon$ of the corresponding substrate and the independently measured (Hall) carrier density $n$ (table 1).

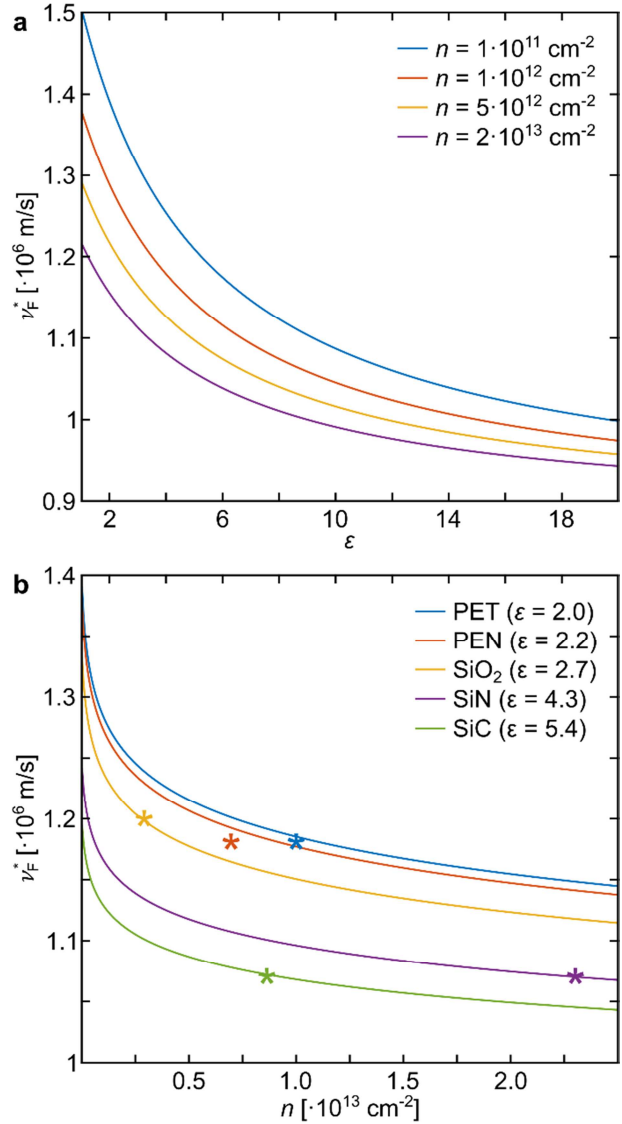

**Figure 1.** Renormalization of Fermi velocity in graphene. (a) Renormalization of Fermi velocity calculated[9] for different static dielectric environments $\varepsilon$ at specific carrier densities $n$. A considerable Fermi velocity renormalization occurs at low dielectric constant $\varepsilon < 3$, with values of $v_F^*$ notably larger than $v_F$. (b) Renormalization of Fermi velocity calculated[9] for different carrier densities for the dielectric environments given by the substrates considered in this study. The extracted renormalized Fermi velocities in our experiments are marked by an asterisk '*' for each of the five substrates considered in this study at the corresponding carrier densities independently obtained from Hall measurements (table 1). Uncertainty in the calculated $v_F^*$ (table 1) is similar or smaller than the size of the asterisk in all cases.

In practical terms, equations 2-4 are a system of three equations with three interdependent unknowns ($n$, $\mu$, $v_F^*$) that can be iteratively solved for each recorded THz spectrum. For completeness, we note that we assume equation 2 (semiclassical approach) to be valid even in the presence of renormalized parameters accounting for weak electron interactions.[25]

## 3. Methods

*3.1 Graphene growth, transfer to flexible substrates and graphene quality on these films*

Graphene on PET was prepared by CVD growth of graphene on copper foil using a low pressure tube furnace system and subsequent transfer onto PET substrates as described in detail previously.[22,27] Commercially available CVD grown graphene (Graphenea) on copper substrate was transferred onto PEN substrates by etching transfer.[28,29] Graphene on $SiO_2$ was fabricated from CVD growth of graphene on Cu(111) foils[30] prepared by contact-free annealing[31] and subsequent bubbling transfer[32]. Electrical parameters $\sigma_{DC}$, $n$ and $\mu$ of graphene in all these substrates measured in this work (via both THz and Hall measurements) show typical values of supported large-scale CVD samples.[15–17,22,33] Specifically, our samples display (figure 3) features already reported in literature for these substrates such as highly homogeneous electrical conductivity of graphene on PET substrates[22] and regions of high mobility of charge carriers in graphene supported on PEN[29] (despite our measurements are undertaken in larger spatial regions and under ambient conditions).

*3.2 Characteristics of substrates used in this work*

PET substrates are 225 µm thick with refractive index of 1.74 at 1 THz. PEN substrates are 125 µm thick with refractive index of 1.83 at 1 THz. The refractive indices and $\varepsilon_s$ for PET, PEN and SiC (refractive index of 3.13) were calculated from THz waveforms of bare substrate relative to air.[34] Our $SiO_2$/Si substrates have a $SiO_2$ thickness of 90 nm with $SiO_2$ having a refractive index of 2.1 at 1 THz.

*3.3 THz-TDS measurements of graphene on arbitrary substrates*

THz-TDS measurements of graphene on polymer foils were performed using two setups: a custom-built broadband air plasma setup[35] and a commercial fiber-coupled spectrometer[36]. For the THz-TDS measurements of graphene on $SiO_2$ (on highly resistive Si) we exclusively used the custom-built broadband air plasma setup and measured at 16 random locations on the sample. All measurements were performed in transmission-mode.

The home-built ultra-broadband THz spectrometer is based on two-color femtosecond air plasma THz generation, which generates pulses as short as few tens of fs, and covers the spectral region from 0.5 up to 30 THz. An air biased coherent detection (ABCD) scheme is employed for smooth, gapless and broadband THz waveform detection.[37,38] Samples are located in the focal plane of the THz beam at normal incidence and can be translated to various sample positions via XYZ linear translation stages.

The air plasma setup can readily measure the conductivity of graphene up to 9 THz – a limit primarily determined by the transmission window of the polymeric substrates. The transmission function $\tilde{T}_{\text{film}}(\omega) = \tilde{E}_{\text{film}}(\omega)/\tilde{E}_{\text{sub}}(\omega)$ where $\tilde{E}_{\text{film}}(\omega)$ and $\tilde{E}_{\text{sub}}(\omega)$ are the Fourier transforms of the THz waveforms transmitted through graphene covered polymer foil and non-graphene covered polymer foil, respectively, is used to calculate the frequency-dependent sheet conductivity of graphene, $\sigma_s(\omega) = \sigma_1 + i\sigma_2$. For the air plasma setup, $\tilde{E}_{\text{film}}(\omega)$ and $\tilde{E}_{\text{sub}}(\omega)$ consists only of a directly transmitted transient and $\sigma_s(\omega)$ can following be determined as:

$$\sigma_s(\omega) = \frac{n_A}{Z_0}\left(\frac{1}{\tilde{T}_{\text{film}}(\omega)}-1\right), \quad (5)$$

where $n_A = n_{\text{sub}}+1$ with substrate refractive index $n_{\text{sub}}$ and $Z_0 = 377\ \Omega$ is the vacuum impedance.[17]

In the commercial setup (Picometrix T-Ray 400), samples were raster scanned in the focal plane of the THz beam at normal incidence to form a map with a spot size at 1 THz of ∼400 µm.[36] For thin polymeric substrates the time difference between internal reflections is smaller than the width of the incident THz pulse in the commercial setup.[22] In such a case, where $\tilde{E}_{\text{film}}(\omega)$ and $\tilde{E}_{\text{sub}}(\omega)$ contain terms from the directly transmitted pulse together with all the following echoes from internal reflections, $\sigma_s(\omega)$ can be determined as:

$$\sigma_s(\omega) = \frac{n_A^2 - n_B^2 e^{-i\delta} + (n_B^2 e^{-i\delta} - n_A^2)\tilde{T}_{\text{film}}(\omega)}{(n_A + n_B^2 e^{-i\delta})Z_0 \tilde{T}_{\text{film}}(\omega)}, \quad (6)$$

where $n_B = n_{\text{sub}}-1$ and $\delta = \omega d n_{\text{sub}}/c$ with substrate thickness $d$.[22] By fitting $\sigma_s(\omega)$ to the Drude model Equation 1 it is possible to extract the DC sheet conductivity $\sigma_{\text{DC}}$ and the carrier scattering time $\tau$ in each measurement point. The carrier density $n$ and mobility $\mu$ can following be determined from $\sigma_{\text{DC}}$ and $\tau$ solving iteratively equations 2-4.

*3.4 Characteristics of substrates used in this work*

Dual configuration Hall effect measurements[39,40] with peripheral electrical contacts (van der Pauw devices) were performed at a constant external magnetic field of 255 mT in a Linkam LN600P stage to determine the Hall carrier density $n_H$ and mobility $\mu_H$ of the samples under study.

The deviation of electrical parameters from THz-TDS and Hall measurements was calculated as (|param$_{\text{THz}}$-param$_H$|)/param$_H$. All THz-TDS and Hall measurements were performed in ambient conditions several months after sample fabrication – this was done in order for samples to stabilize since the measured electrical properties of graphene are extremely sensitive to surrounding conditions.[41,42]

**4. Results**

*4.1 THz-TDS of graphene films on arbitrary thin substrates*

Large-scale (cm size) graphene films were prepared by chemical vapor deposition (CVD) on copper foil prior to its transfer onto flexible polymer substrates (PEN, PET). Notably, $\varepsilon_s$ (and thereby the effective dielectric constant $\varepsilon$) is low in both substrates ($\varepsilon_s \sim 3$). Furthermore, we have prepared one more graphene sample on highly resistive silicon substrate with a 90 nm thermal oxide ($SiO_2$) on top. Since $SiO_2$ also has a low $\varepsilon_s$ (~4), the extraction of electrical parameters of graphene films via THz-TDS on this rigid and commonly utilized substrate[3,27,43–45] demonstrates the versatility of these results for arbitrary substrates. Additional characteristics of each of the substrates can be found in Supplementary Table S1.

Prior to the THz-TDS characterization, Hall measurements were performed to independently determine the Hall carrier density $n_H$ and Hall mobility $\mu_H$ of the samples under study. These independent values ($n_H$, $\mu_H$) will be compared with the median carrier density and mobility values extracted from maps of THz-TDS spectra, as routinely done in literature.[15,16,33]

THz-TDS measurements of graphene on thin polymer films and $SiO_2$/Si were performed in the present work by using two different set-ups: a custom-built, ultra-broadband (range ~2-10 THz) air plasma setup[35] and a commercial fiber-coupled broadband spectrometer (range ~0.5-1.5 THz)[36]. The commercial system allows the electrical properties of large-scale CVD graphene to be mapped quickly and precisely.[15,17] In contrast, the air plasma setup with its much broader frequency range allows us to verify the values extracted for $\sigma_{DC}$ and $\tau$ from the commercial setup due to the greater accuracy of the fits to the Drude model (equation 1). We point out that that this consistency test is desirable when using thin (~100 µm) polymeric films. The timing between internal reflections in these substrates (~1-2 ps) is commonly shorter than the width of the time-domain waveforms in the commercial setup (see figure 2(a)), and the frequency-dependent sheet conductivity $\sigma_s(\omega)$ must be calculated by using a complicated fitting expression accounting for internally reflected transients[22] since THz transients from direct transmission and internal reflections are overlapping, and thus occurring within the same waveform. In contrast, $\sigma_s(\omega)$ can be extracted using a simpler formalism in the broadband setup, covering only a directly transmitted transient since the width of the time-domain waveforms in this setup is shorter than the intervals between internal reflections as shown in figure 2(a).

Figure 2(a) shows the measured THz time-domain waveforms and their corresponding Fourier transform from both commercial and air plasma set-ups after transmission through graphene on a thin flexible PEN substrate. Examples of $\sigma_s(\omega)$ measured from both setups for graphene on PEN are shown in figure 2(b), together with their corresponding fits to the Drude model. This figure demonstrates the consistent values ($\sigma_{DC}$, $\tau$) extracted from both commercial and broadband air-plasma systems showing agreement better than ~3%. A similar comparison was done for graphene on PET as shown in Supplementary figure S1.

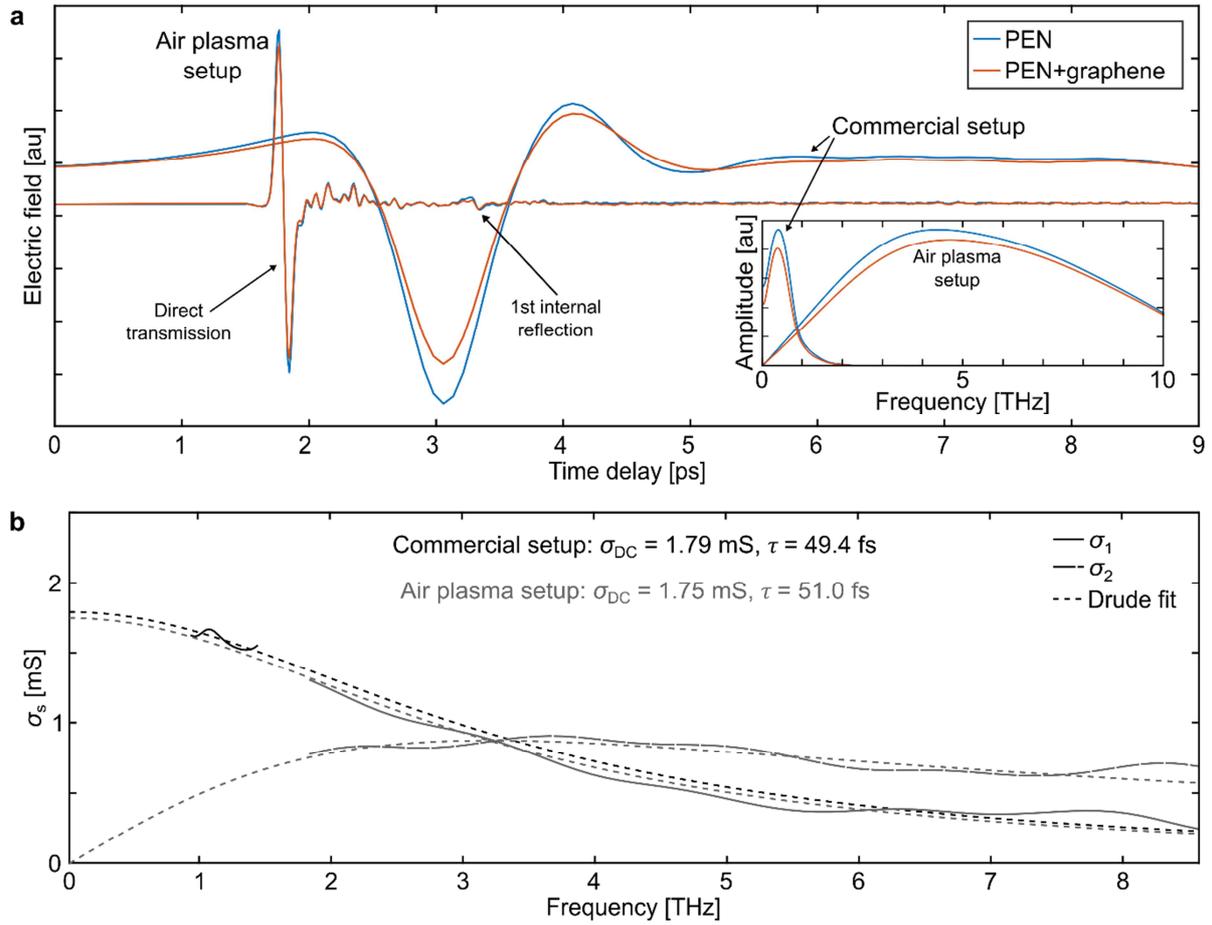

**Figure 2**. THz-TDS of graphene on PEN. (a) Waveforms of THz pulses after transmission through PEN substrate without and with graphene. Inset shows the Fourier transform of the time-domain waveforms. (b) Comparison of sheet conductivity extracted from THz-TDS measurements in the same region of a sample of graphene on PEN using both commercial and air plasma-based set-ups, respectively. Dashed lines are fits to the experimental data by the Drude model.

*4.2 Extracted electronic ($v_F^*$) and electrical (n, μ) parameters of graphene films on arbitrary thin substrates via THz-TDS*

With these values ($\sigma_{DC}$, $\tau$), we can then proceed to calculate $n$, $\mu$ and $v_F^*$ of graphene on PEN, PET and SiO$_2$ (Supplementary figure S2) by iteratively solving Equations 2-4. Maps and histograms of the electrical parameters ($\sigma_{DC}$, $\tau$, $n$ and $\mu$) of graphene on PEN and on PET are shown in figure 3 and Supplementary figures S3 and S4. The overall conductivity in these samples is homogeneous (figure 3(a,b) (insets) and Supplementary figures S3(e) and S4(e)). Histograms of $n$ and $\mu$ from both samples are shown in figure 3, having values that are typical for large-scale CVD graphene supported on rigid substrates[15–17,22,33]. The median values of $n$ and $\mu$ from the THz-TDS measurements and the corresponding $n_H$ and $\mu_H$ from Hall measurements are also highlighted in the figure (see also Table 1), and show a reasonable quantitative agreement for both substrates (detailed analysis in the next subsection). Moreover, since $v_F^*$ depends on n in each measurement pixel, we can also map the variation in $v_F^*$ across a sample. Maps of $v_F^*$ for graphene on PEN and PET are shown in figure 4 together with their corresponding histogram. First, the averaged values of $v_F^*$ is ~$1.18 \cdot 10^6$ m/s for both PEN and PET; confirming an appreciable variation with respect to $v_F$ and/or the commonly used value ~$1 \cdot 10^6$ m/s.[46] We emphasize that this occurs even at large carrier densities ~$1 \cdot 10^{13}$ cm$^{-2}$ and ~$5 \cdot 10^{12}$ cm$^{-2}$ measured for the PET and PEN cases, respectively, due to the low permittivity of these substrates. Furthermore, $v_F^*$ maps are very homogeneous in both cases: histograms have standard deviations on fitted normal distributions below 2 % for both polymeric substrates. The latter is expected due to two factors: *i*) standard deviations of *n* are also relatively small in both substrates and *ii*) the $\ln(\Lambda/\sqrt{\pi n})$ dependence of $v_F^*$ (equation 4). Additional considerations regarding the accuracy of the estimated $v_F^*$ by the here presented method can be found below.

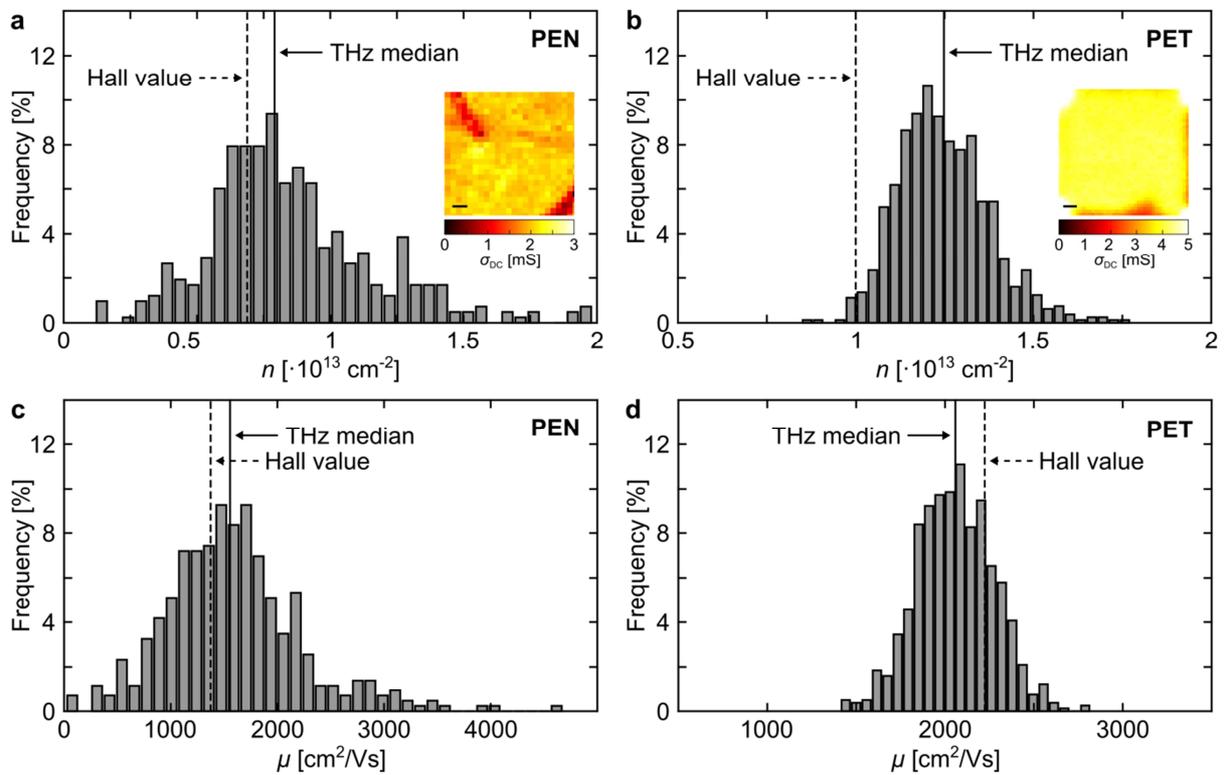

**Figure 3**. Histograms of electrical parameters. Histograms of (a,b) carrier density and (c,d) mobility from THz-TDS measurements of graphene on (a,c) PEN and (b,d) PET. Dashed lines show value measured from Hall measurement on the same sample. Insets in (a,b) show the DC conductivity map from the sample (see also Supplementary figures S3 and S4). Scale bars in insets are 1 mm.

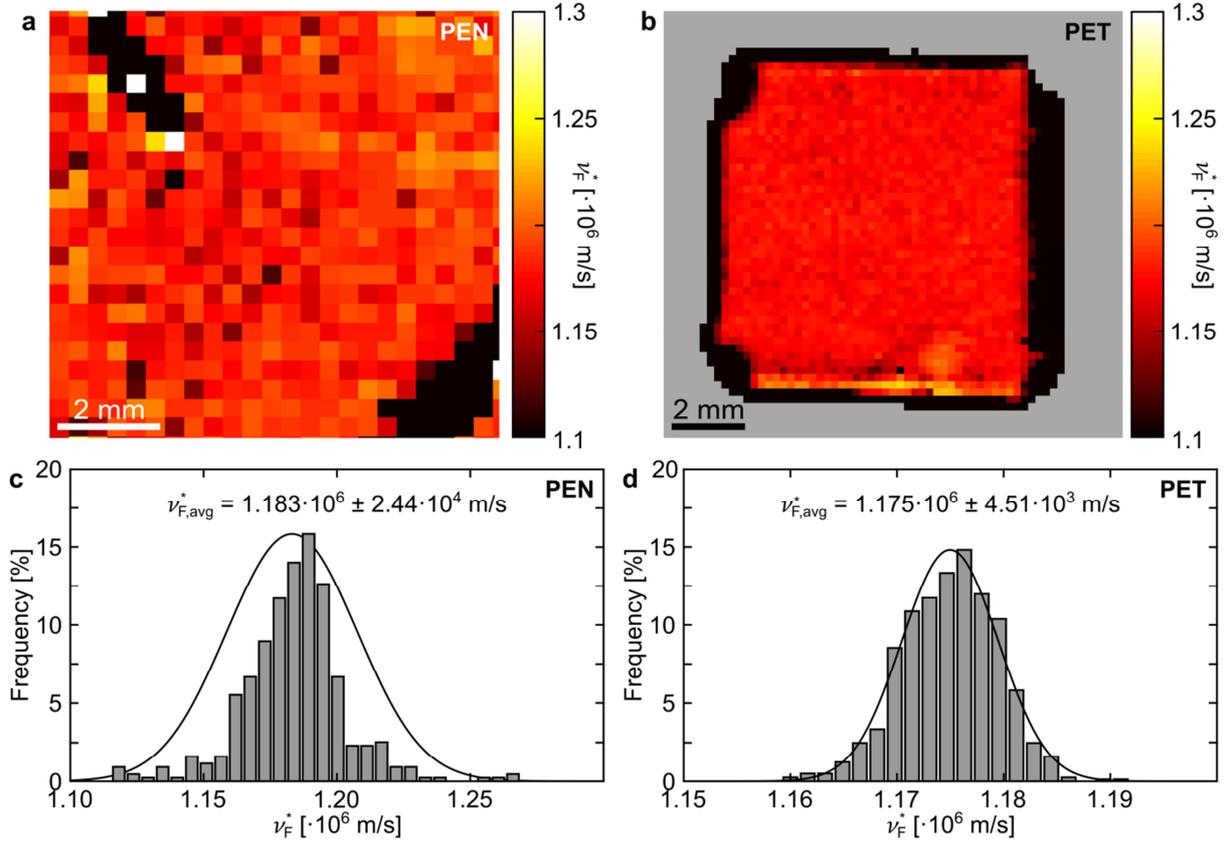

**Figure 4.** Maps and histograms of Fermi velocity. (a,b) Maps and (c,d) corresponding histograms of Fermi velocity for the samples of graphene on PEN and PET shown in figure 3. The black lines in (c,d) show fitted normal distributions.

A complete list of all median values of $n$, $\mu$ and $v_F^*$ extracted from THz-TDS and $n_H$ and $\mu_H$ obtained from Hall measurements for all of the substrates assessed in this study (PEN, PET, $SiO_2$, SiN and SiC) can be found in table 1. In addition, this table shows deviations between $n$, $\mu$ and $n_H$, $\mu_H$ and the estimated (Supplementary Material - Methods) relative error (uncertainty) of the measured $n$, $\mu$ and $v_F^*$ ($\Delta v_F^*/v_F^*$) in all substrates.

|  | $n_H$ ($\cdot 10^{13}$ $cm^{-2}$) | $\mu_H$ ($cm^2/Vs$) | $n$ ($\cdot 10^{13}$ $cm^{-2}$) | $\mu$ ($cm^2/Vs$) | $\Delta n/n_H$ (%) | $\Delta \mu/\mu_H$ (%) | $v_F^*$ ($\cdot 10^6$ m/s) | $\Delta v_F^*/v_F^*$ (%) |
|---|---|---|---|---|---|---|---|---|
| PEN | 0.69 | 1398 | 0.79 | 1548 | 15 | 13 | 1.18 | 0.55 |
| PET | 1.00 | 2214 | 1.23 | 2057 | 23 | 7 | 1.18 | 0.86 |
| $SiO_2$ | 0.29 | 1705 | 0.29 | 1728 | 2 | 2 | 1.20 | 0.07 |
| SiN[15] | 2.30 | 900 | 2.07 | 969 | 10 | 8 | 1.07 | 0.29 |
| SiC[16] | 0.86 | 3413 | 0.95 | 3956 | 11 | 16 | 1.07 | 0.28 |

**Table 1.** Extracted electrical and electronic parameters. Comparison of $n$, $\mu$, $v_F^*$, relative error of $n$, $\Delta n/n_H$ (where $\Delta n = |n-n_H|$), relative error of $\mu$, $\Delta \mu/\mu_H$ (where $\Delta \mu = |\mu-\mu_H|$) and uncertainty in the extracted $v_F^*$ ($\Delta v_F^*/v_F^*$). Values $n$, $\mu$, $v_F^*$ are extracted by using equations 2-4, with $\varepsilon_s$, $\sigma_{DC}$ and $\tau$ for each substrate given in Supplementary table S1.

## 5. Discussion

Table 1 shows the extracted $v_F^*$ of graphene on the five different considered substrates. Notably, in some substrates, $v_F^*$ reaches values 20% larger than the commonly used Fermi velocity value of $1 \cdot 10^6$ m/s.[46] The overall variation in the values of $v_F^*$ for all considered substrates is ~11% (largest value $1.2 \cdot 10^6$ m/s, lowest value $1.07 \cdot 10^6$ m/s). In addition, one can see that the renormalized Fermi velocity $v_F^*$ is larger in graphene on substrates with lower permittivity < 5 (PEN, PET and SiO$_2$) with respect to graphene on substrates with higher permittivity (SiN and SiC). This is expected and agrees with theory (see section before results and figure 1(b)). Importantly, we note that the estimated uncertainty in $v_F^*$, $\Delta v_F^*/v_F^*$, is < 0.9% in all cases, an order of magnitude smaller than the aforementioned ~11% measured variation in the values of $v_F^*$ in all substrates. The latter demonstrates the ability of the here proposed THz-TDS technique to probe $v_F^*$ in graphene even at large densities $n > 1 \cdot 10^{12}$ cm$^{-2}$.

Extending this more exhaustive analysis to other parameters, we first point out that deviations between all ten values of $n$, $\mu$ and $n_H$ and $\mu_H$ for all substrates are commonly low, below 15-16%. This agreement highlights the fact that THz-TDS is also a viable method for characterizing the electrical properties ($\sigma_{DC}$, $\tau$, $n$, $\mu$) of graphene on arbitrary substrates. Values $\mu$, $n$ extracted in graphene on (flexible) substrates with low permittivity, PEN and PET, would exhibit much larger and systematic errors (w.r.t. the independently measured values $\mu_H$, $n_H$) if a constant Fermi velocity value of $v_F = 1 \cdot 10^6$ m/s is assumed instead of $v_F^*$. Specifically, errors using a fixed $v_F = 1 \cdot 10^6$ m/s would be ×1.6 for $\mu$, ×3.9 for $n$ and ×4.7 for $\mu$, ×3.0 for $n$ times larger for PEN and PET, respectively. We further note that, when using $v_F^*$, only one value shows a deviation slightly larger than 15% between THz-TDS and Hall measurements in these two substrates (~23 % observed for $n$ in PET), and this particular case can be accounted for by residual (device-scale) inhomogeneities[39] present in graphene on this polymer.

Moreover, we point out the excellent agreement (deviations below 2% in both $n$ and $\mu$) obtained between THz-TDS and Hall measurements undertaken for graphene on SiO$_2$ (Table 1); a commonly utilized (rigid) low permittivity substrate, where the transfer of defect-free and homogeneous graphene is well-optimized.[27,47–49] The verification of the here presented technique for SiO$_2$ substrates is highly relevant since this is the standard substrate used when characterizing the properties of graphene via additional measurement techniques such as field effect transistor[50] and Hall[39] measurements, Raman spectroscopy[51–53] and optical microscopy[54]. More importantly, such outstanding agreement further emphasizes the non-negligible impact of electron-electron interactions even in doped graphene: notable deviations have been reported[55] in similar measurements on SiO$_2$ when neglecting these many-body effects. In our case, systematic deviations (> 40%) would exist between both $n$, $\mu$ and $n_H$, $\mu_H$ if a fixed Fermi velocity value of $1 \cdot 10^6$ m/s is assumed for graphene on $SiO_2$, deviations which are $> \times 20$ times larger than the ones extracted when using $v_F^*$.

Finally, Table 1 also shows and assess the consistency of our method with results already reported in literature for two additional (rigid) substrates with larger permittivity (SiN[15], SiC[16]). By renormalizing the Fermi velocity as proposed in this study, we demonstrate an even better overall agreement between THz-TDS and Hall measurements compared to the previously reported values (i.e. deviations are reduced by 2-3 times even in these two substrates with larger permittivity). Specifically, by accounting for $v_F^*$, the deviation between THz-TDS and Hall measurements decreases from 36% to 10% for carrier density and from 25% to 8% for mobility in the SiN substrate case[15]. For the SiC substrate[16] the deviation is similar (9% to 11%) for carrier density values and decreases from 28% to 16% for mobility values.

We remark that Fermi velocity renormalization effects described here only influence the extracted values of $n$ and $\mu$ from THz-TDS measurements (see section 2). As such, previously measured and reported values of $\sigma_{DC}$ and $\tau$ in the literature [15,16,21-23,36,42] are accurate, i.e. not affected by this correction.

## 6. Conclusion

In conclusion, we have demonstrated THz-TDS to be an accurate, rapid and scalable method to probe the renormalized Fermi velocity $v_F^*$ in graphene. This additionally allows to quantitatively obtain all electrical parameters (conductivity $\sigma_{DC}$, carrier density $n$ and carrier mobility $\mu$) of graphene placed on arbitrary substrates. Moreover, we have demonstrated that graphene charge carriers on films with low relative permittivity (< 5) such as flexible polymeric substrates are notably subjected to electron-electron interactions, due to both dielectric screening and self-screening mechanisms, [1,7–9] even at relatively large carrier densities $> 1 \cdot 10^{12}$ $cm^{-2}$. From a technological point of view, these results demonstrate the accuracy and versatility of contactless THz-TDS to quantify the electrical properties of graphene within different surrounding environments, enabling its utilization to the many applications[11–14] and production scenarios[27,56] involving large-scale graphene films on arbitrary insulating and/or flexible substrates, or encapsulated graphene. The here presented results are extendable to any Dirac material.[9]

**Acknowledgements**

Authors declare no competing interests.

We acknowledge stimulating discussions with D.H. Petersen and E. Díez.

This work was supported by the Danish National Research Foundation (DNRF) Center for Nanostructured Graphene (DNRF103); EU Graphene Flagship Core 2 (785219); Villum Fonden (00023215); Natural Science Foundation of the People's Republic of China (NSFC) (51402291, 51702315); Chongqing Research Program of Basic Research: Frontier Technology (cstc2015jcyjA50018); Institute for Basic Science (IBS-R019-D1); Vinnova (2019-02878).

# Supplementary material

# Fermi Velocity Renormalization in Graphene Probed by Terahertz Time-Domain Spectroscopy

Patrick R. Whelan,[1,2,*] Qian Shen,[3,4] Binbin Zhou,[3] I. G. Serrano,[5] M. Venkata-Kamalakar,[5] David M. A. Mackenzie,[6] Jie Ji,[1,2] Deping Huang,[7] Haofei Shi,[7] Da Luo,[8] Meihui Wang,[9] Rodney S. Ruoff,[8,9,10,11] Antti-Pekka Jauho[1,2], Peter U. Jepsen,[2,3] Peter Bøggild[1,2] and José M. Caridad[1,2,*]

[1]*DTU Physics, Technical University of Denmark, Ørsteds Plads 345C, DK-2800 Kongens Lyngby, Denmark*
[2]*Center for Nanostructured Graphene (CNG), Technical University of Denmark, Ørsteds Plads 345C, DK-2800 Kongens Lyngby, Denmark*
[3]*DTU Fotonik, Technical University of Denmark, Ørsteds Plads 343, DK-2800 Kongens Lyngby, Denmark*
[4]*School of Information Engineering, Nanchang University, Nanchang 330031, P. R. China*
[5]*Department of Physics and Astronomy, Uppsala University, Box 516, SE 751 20, Uppsala, Sweden*
[6]*Department of Electronics and Nanoengineering, Aalto University, P.O. Box 13500, FI-00076 Aalto, Finland*
[7]*Chongqing Institute of Green and Intelligent Technology, Chinese Academy of Sciences, 266 Fang Zheng Ave., Chongqing 400714, P. R. China*
[8]*Center for Multidimensional Carbon Materials (CMCM), Institute for Basic Science (IBS), Ulsan 44919, Republic of Korea*
[9]*Department of Chemistry,* [10]*School of Materials Science and Engineering, and* [11]*School of Energy and Chemical Engineering, Ulsan National Institute of Science and Technology (UNIST), Ulsan 44919, Republic of Korea*

Emails: patwhe@dtu.dk, jcar@dtu.dk

## S1. Supplementary methods

Following Equation 4 in the main text and assuming no error in $\varepsilon$ (i.e. $\alpha$), the absolute error of $v_F^*$ ($\Delta v_F^*$) is given by:

$$\Delta v_F^* = \left|\frac{\partial v_F^*}{\partial n}\right| \Delta n \quad (7)$$

where $\Delta n$ is the absolute error of the carrier density $n$, and

$$\left|\frac{\partial v_F^*}{\partial n}\right| = \frac{v_F C(\alpha)\alpha}{2n}. \quad (8)$$

In other words

$$\Delta v_F^* = \frac{v_F C(\alpha)\alpha}{2}\frac{\Delta n}{n} \quad (9)$$

where $\Delta n/n$ is the relative error of the carrier density $n$. Here, we estimate this error from the independently extracted carrier density via Hall measurements $n_H$ (i.e. $\Delta n/n = |n-n_H|/n_H$ (see Table 1, main text).

We note that the relative error of $v_F^*$ ($\Delta v_F^*/v_F^*$) extracted for all substrates is below 0.9% (see Table 1, main text).

## S2. Supplementary figures

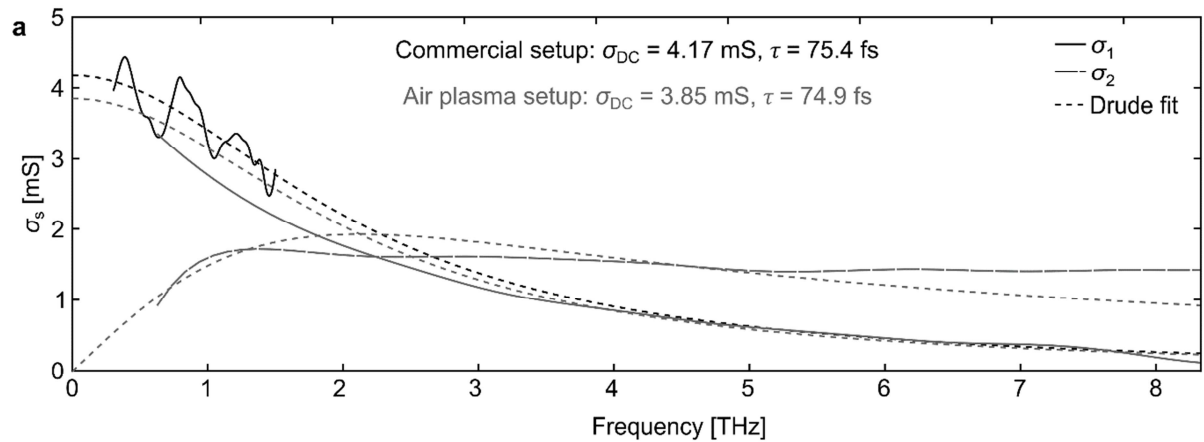

**Supplementary figure S1.** THz-TDS conductivity of graphene on PET. (a) Comparison of sheet conductivity extracted from THz-TDS measurements in the same region of a sample of graphene on PET using a commercial and air plasma-based setup, respectively. Dashed lines show fits to the Drude model.

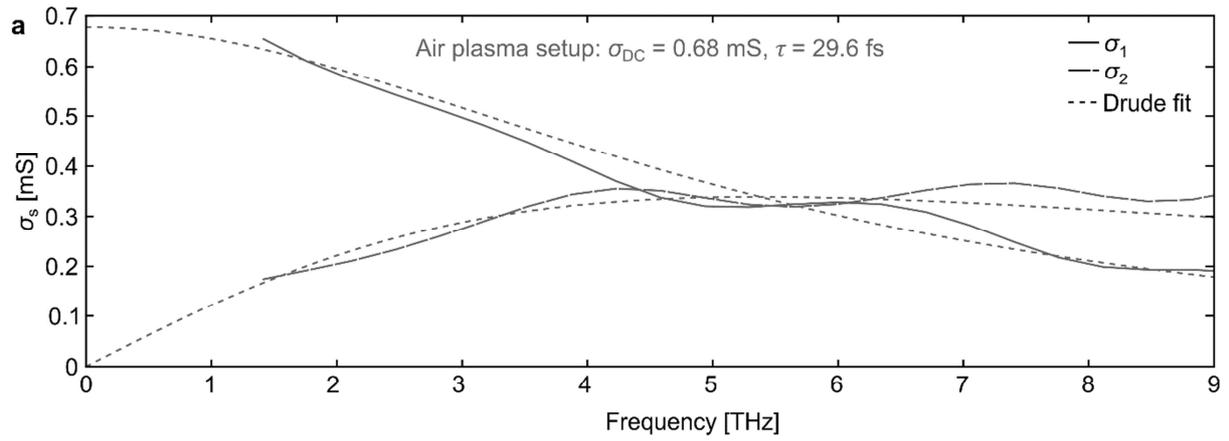

**Supplementary figure S2.** THz-TDS conductivity of graphene on $SiO_2$. (a) Example of sheet conductivity extracted from THz-TDS measurements of a sample of graphene on $SiO_2$ using an air plasma-based setup. Dashed lines show fits to the Drude model.

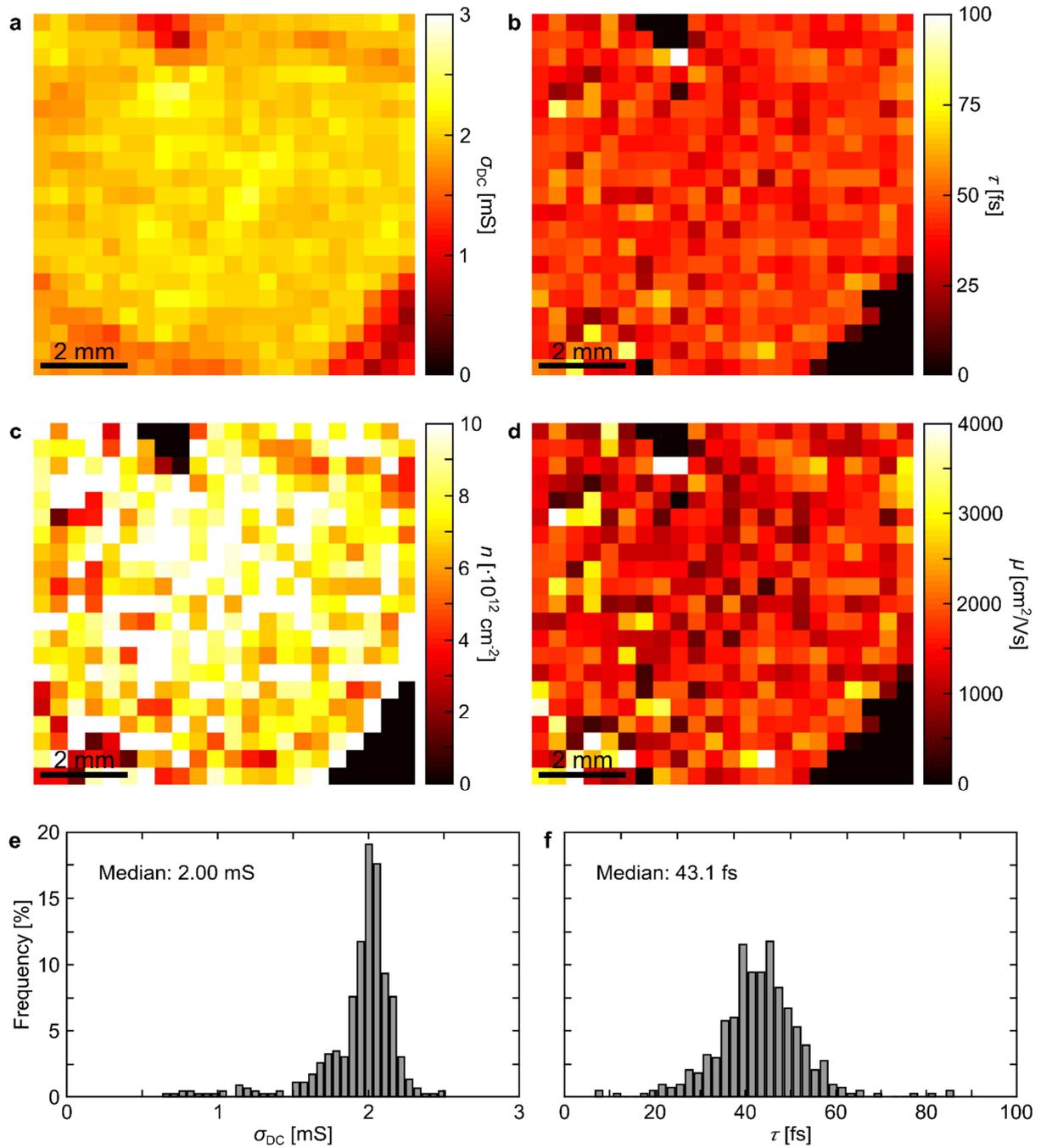

**Supplementary figure S3.** THz-TDS data for graphene on PEN. (a-d) Maps of THz-TDS (a) DC sheet conductivity, (b) scattering time, (c) carrier density, and (d) mobility for the sample of graphene on PEN shown in figure 3. (e,f) Histograms of extracted (e) DC sheet conductivity and (f) scattering time.

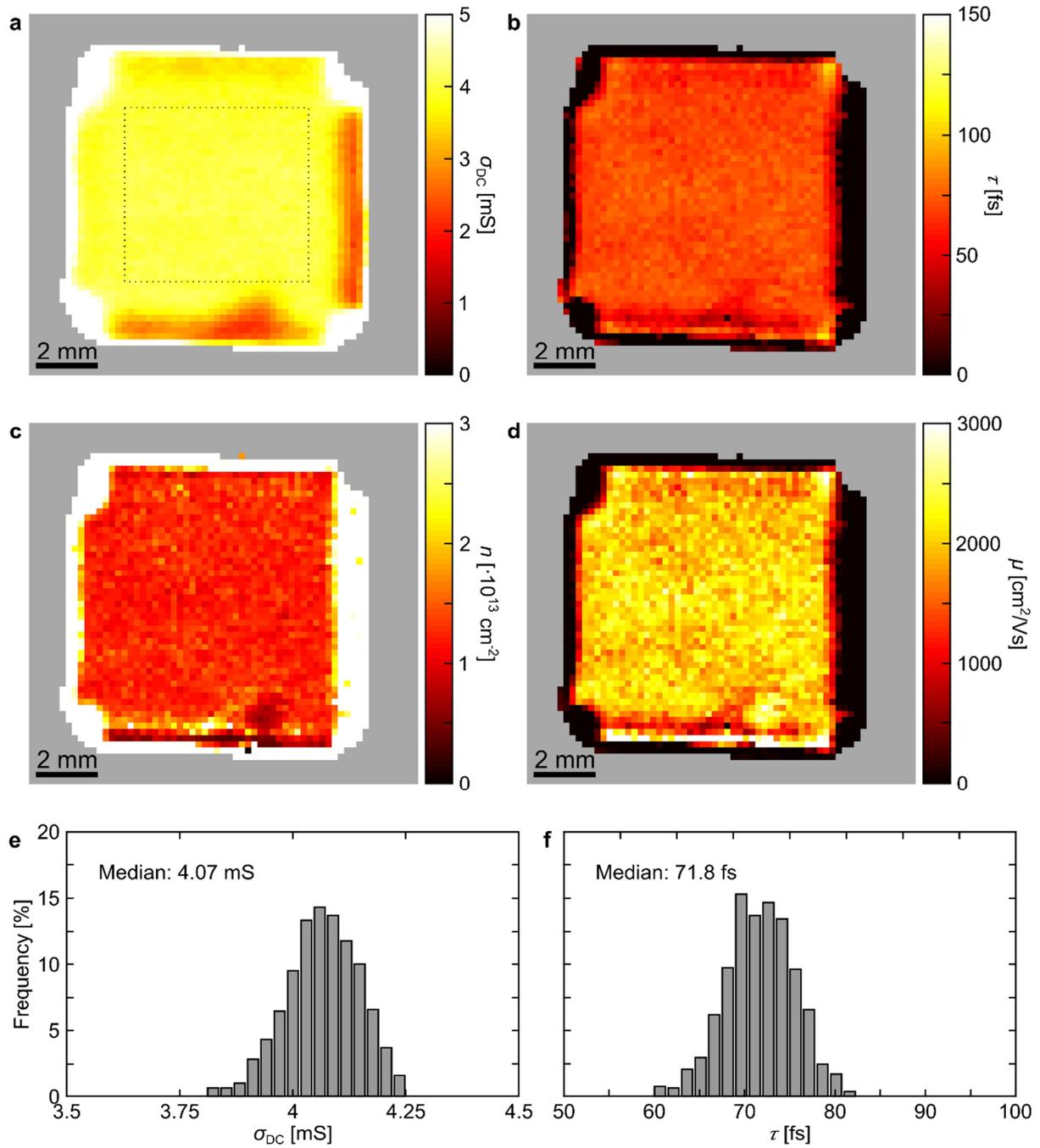

**Supplementary figure S4.** THz-TDS data for graphene on PET. (a-d) Maps of THz-TDS (a) DC sheet conductivity, (b) scattering time, (c) carrier density, and (d) mobility for the sample of graphene on PEN shown in figure 3. (e,f) Histograms of extracted (e) DC sheet conductivity and (f) scattering time. The histogram values are taken from the data points inside the square highlighted in (a).

## S3. Supplementary table

|       | $\varepsilon_s$ | $\sigma_{DC}$ (mS) | $\tau$ (fs) |
|-------|-----|-----|-------|
| PEN   | 3.3 | 2.00 | 43.1 |
| PET   | 3.0 | 4.07 | 71.8 |
| SiO$_2$ | 4.4 | 0.79 | 28.4 |
| SiN[1] | 7.5 | 3.22 | 48.0 |
| SiC[2] | 9.8 | 6.05 | 133.4 |

**Supplementary table S1**. Comparison of $\varepsilon_s$, $\sigma_{DC}$, and $\tau$ extracted from THz-TDS measurements. The values for $\varepsilon_s$ are extracted from THz-TDS measurements undertaken for PET, PEN and SiC (see Methods in main), while we have used literature values at 1 THz for SiO$_2$[3] and SiN[4].

## Supplementary References